\documentclass{article}
\usepackage{arxiv}
\usepackage{graphicx} 
\usepackage{subcaption}
\usepackage{float}
\usepackage{amsmath}
\usepackage{amssymb}
\usepackage{xcolor}
\usepackage[breakable, theorems, skins]{tcolorbox}
\usepackage{makecell}
\usepackage{todonotes}
\usepackage[sorting=none]{biblatex}
\usepackage{setspace}
\usepackage[switch]{lineno}
\usepackage{hyperref}
\usepackage{hyperref}

\title{Optimizing infectious disease mitigation under dynamic conditions}

\date{}

\usepackage{authblk}

\author[1,2]{Laura M\"uller}
\author[1,3]{Fabio Sartori}
\author[1,2]{Jonas Dehning}
\author[4,5]{Maximilian F. Eggl}
\author[1,2*]{Viola Priesemann}

\affil[1]{Max Planck Institute for Dynamics and Self-Organization, G\"ottingen, Germany.}
\affil[2]{Institute for the Dynamics of Complex Systems, University of G\"ottingen, G\"ottingen, Germany.}
\affil[3]{Karlsruhe Institute of Technology, Karlsruhe, Germany.}
\affil[4]{Institute of Neuroscience, CSIC-UMH, Alicante, Spain.}
\affil[5]{Institute of Experimental Epileptology and Cognition Research,
University of Bonn Medical Center, Bonn, Germany.}

\affil[ ]{* Corresponding Author: Viola Priesemann (viola.priesemann@ds.mpg.de)}

\begin{document}
\maketitle

\begin{abstract}
Mitigation measures are essential for controlling the spread of infectious diseases during pandemics and epidemics, but they impose considerable societal, individual, and economic costs. We developed a general optimization framework to balance costs related to infection and to mitigation. Optimizing the trade-off between mitigation and infection cost, we identified three novel, surprising effects:
First, assuming a constant reproduction number $R_0$, the optimal response to an infectious disease requires either strict mitigation or none at all, depending on disease severity, but never does one find an intermediate mitigation level to be optimal.
Second, under seasonal variations, optimal mitigation is stricter during winter. Interestingly, a single wave of infections still arises in spring with 3 months delay to the seasonal peak of infectivity, replacing the autumn/winter waves known for classical influenza.
Third, during steady vaccination campaigns, even optimal mitigation can result in transient infection waves. Finally, we quantify the cost of delayed mitigation onset and show that even short delays can substantially increase total costs --- if the disease is severe.
Overall, our framework is easily applicable to general and complex settings and thereby presents a versatile tool to explore optimal mitigation strategies for endemic and pandemic infectious disease.
\end{abstract}


\clearpage
\section{Introduction}

During pandemics like COVID-19, societies face the challenge of implementing measures to mitigate disease spread. These measures, however, come at considerable cost, including economic repercussions \cite{Deb2022}, as well as social and psychological burdens \cite{MARROQUIN2020}. Still, they can be necessary to reduce the number of new infections, prevent healthcare systems from becoming overwhelmed, and minimize sick leave and preventable deaths. This trade-off raises the fundamental question in pandemic response: What is the optimal mitigation stringency, and how does that change over time and under different conditions? Determining the optimal, time-dependent level of mitigation requires quantifying and balancing infection costs against mitigation costs. This challenge becomes especially complex in long-term pandemic scenarios.

While the optimization of vaccination strategies has been extensively studied \cite{Longini1978, Lopes2022review, Matrajt2021, Han2021}, vaccines are often unavailable during the early phase of a pandemic, if at all. Our study thus centers on mitigation relying on non-pharmaceutical interventions (NPIs) such as social distancing, lockdowns, and travel restrictions, including both voluntary and mandatory measures.

Several studies explore mitigation strategies by comparing a limited set of fixed scenarios \cite{contreras2021, brauner2021inferring, oraby2021modeling, gemmetto2014mitigation, dehning2020inferring}. Others use systematic optimization methods based on optimal control theory \cite{gavish2025optimal, Greenhalgh1988, Shirin2021, Albi2021, Macalisang2020, Godara2021} or model predictive control \cite{selley2015, Peni2020, Martins2023}. For the cases studied, these algorithms typically yield very similar or even identical results. More recently, machine learning algorithms, such as reinforcement learning, have been used in addition to the classical approaches \cite{Ohi2020, arango2020, Hwang2022, Libin2021}. Each of these strategies has its strengths and limitations: Classical optimization methods offer interpretability, while machine learning approaches may provide more flexibility and adaptability, though their optimization processes can be less transparent.

For non-pharmaceutical interventions, a constant level of mitigation measures is optimal during most of the pandemic if one assumes a constant basic reproduction number \cite{Shirin2021}. If, however, the goals, disease parameters, boundary conditions, or cost functions are more complex, studies reach different conclusions. For instance, a strong lockdown followed by gradual relaxation is optimal in many settings influenced by real-world scenarios, whereas premature relaxation can result in large secondary waves \cite{Perkins2020, contreras2021,Martins2023}. If maintained at low numbers, infections can be kept under control with only mild measures complemented by test-trace-and-isolate strategies \cite{contreras2021, Contreras2021-2, kretzschmar2021isolation}. Additionally, early intervention at the beginning of a pandemic is crucial to prevent large outbreaks and reduce costs \cite{Peni2020, Albi2021, Perkins2020}.

Despite this progress, important gaps remain. Most existing studies assume a constant or only simply varying basic reproduction number. One study included a linearly increasing reproduction number, for which optimal control measures also need to intensify correspondingly \cite{Shirin2021}. Similarly, mitigation efforts can be relaxed when reproduction numbers decline due to vaccination \cite{Shirin2021, Bauer2021}. However, even those scenarios where the reproduction number is varied are still relatively simple. Additionally, most optimizations have been performed using variants of the SIR model without waning immunity, or by assuming that vaccination is distributed sufficiently quickly and is long-lasting. However, for some infectious pathogens, no effective or long-lasting vaccines exist (e.g., HIV, Epstein-Barr virus, SARS-CoV-2, and certain bacterial infections like gonorrhea). For these, long-term optimization of mitigation strategies might become necessary. This requires developing a versatile optimization framework for mitigation in long-term dynamical scenarios.

In this work, we present a comprehensive framework for determining optimal mitigation strategies by minimizing the total cost of a pandemic, consisting of mitigation and infection costs. We introduce general cost functions whose shapes follow basic assumptions about decision-making. 
Our framework's flexibility enables the exploration of a wide variety of scenarios. We demonstrate this versatility by first investigating the impact of seasonal fluctuations in the reproduction number on the optimal mitigation strategy and comparing it to alternative strategies. Next, we simulate vaccination campaigns to assess how increased population immunity through vaccination can allow for a decrease in mitigation measures. Finally, we analyze the consequences of delayed mitigation efforts at the start of a new pandemic. Our analysis quantifies key outcomes, including the total number of infections, the required intensity of measures, and the overall costs associated with disease management.

\section{Methods}

\begin{figure*}
    \centering
    \includegraphics{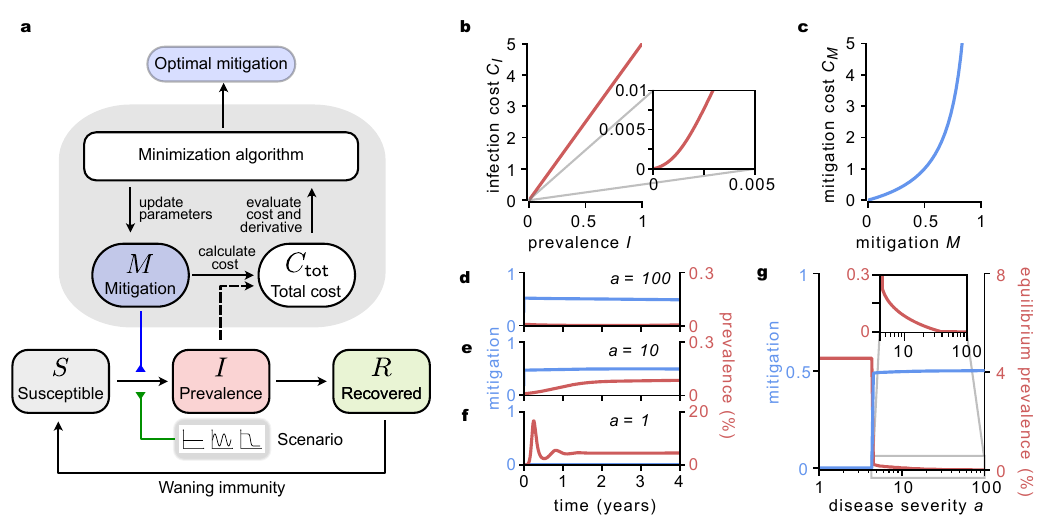}
    \caption{
    \textbf{Overview of the optimization framework --- For a constant reproduction number, optimal mitigation is ``all-or-nothing".} \textbf{(a)} The SIRS model consists of three compartments: \textit{Susceptible (S)}, \textit{Infectious (I)}, and \textit{Recovered (R)}. The flow from \textit{Susceptible} to \textit{Infectious} is determined by the effective reproduction number $R_{\textsf{eff}}$, which is dependent on the scenario (for example, seasonality) and the mitigation $M$. For each simulation, the total cost $C_\text{tot}$ is calculated. By using the total cost and its gradient, the minimization algorithm (L-BFGS) updates the mitigation parameters and eventually returns the optimal mitigation when a local minimum of the cost function is found. \textbf{(b)} We assume that the infection cost increases with the prevalence. \textbf{(c)} We assume that the mitigation cost grows supra-linearly with mitigation strength. The function diverges because it is impossible to realize complete eradication, i.e. when mitigation $M=1$. \textbf{(d-f)} For any disease severity $a$, which defines the ratio between infection and mitigation cost, optimal mitigation converges to a steady state: The optimal mitigation is high for high and intermediate disease severity (d, e), whereas the prevalence is high, or at an intermediate level, for low and intermediate disease severity respectively (f, e). We will be mostly interested in the intermediate ($a$=10) regime, where both variables are non-vanishing. 
    \textbf{(g)} The transition between the low and the high optimal mitigation regime is abrupt: 
    At a critical value of $a$, mitigation jumps from 0 to approximately 0.5, resulting in a low, approximately constant prevalence. 
    }
    \label{fig:model}
\end{figure*}

Our approach is to define a cost function (I) that depends on the disease dynamics (II) and mitigation measures, and optimize (III) the resulting system to minimize the overall integrated costs. In more detail:

\paragraph{(I) Cost functions.} The total cost of infectious disease mitigation comprises two components: the cost associated with infections and the cost of mitigation. 

We assume that the infection cost increases approximately quadratically with the prevalence $I$ up to 0.1\,\% of the population and transitions to a linear increase beyond this threshold, which can be interpreted as a hospitalization threshold (Fig.~\ref{fig:model}b; we also consider other cost functions, see Fig.~S\ref{SI_fig:costplot}):

\begin{equation}
\label{eq:CI}
\begin{split}
    C_{I}(t)\, =&\frac{I(t)}{v}+v\left( I\left(t\right)-0.001\right)\cdot\\ &\left(1-\left[1+\exp(1000(I\left(t\right)-0.001))\right]^{-1}\right)+w, 
\end{split}
\end{equation}
where $v = 4.7913$ and $w = 0.0013$ scale the transition from quadratic to linear growth of the infection cost. 
This pattern reflects the escalating costs associated with rising infection rates, which we assume become linear (i.e.~simply proportional to the number of infectious persons) once healthcare systems reach their hospitalization capacity.

The mitigation cost follows a diverging function, where $M(t) \in [0,1)$ represents the mitigation level (Fig.~\ref{fig:model}c).
\begin{equation}
    C_{M}(t) = \frac{M\left(t\right)}{1-M\left(t\right)}. \label{eq:CM}
\end{equation}
As in any rational mitigation scheme, one first uses low-cost measures (e.g., hygiene and mask recommendations). If stronger mitigation becomes necessary, costs rise supra-linearly due to increasing costs when further transmission is reduced. As $M$ approaches 1, the cost diverges, reflecting the practical impossibility of completely stopping transmission due to unavoidable transmission routes (e.g., healthcare worker-patient interactions; or non-compliance with social distancing).

The total cost, $C_{\textsf{tot}}$, is the sum of infection and mitigation costs and is integrated over the entire simulation period (Eq.~\eqref{eq:Ctot}). To account for the relative importance of these costs, we introduce a disease severity parameter, $a$, which scales the infection costs relative to mitigation costs. Note that $a$ is a characteristic of a disease, but also partially subjective as it encodes normative or cultural judgments about how infection-related harms should be weighed against the societal and economic costs of mitigation.

\begin{equation}
    C_{\textsf{tot}}= \int_0^T \left(aC_I(t)+ \,C_M(t)\right) dt. \label{eq:Ctot}
\end{equation}

To facilitate comparison across scenarios with differing durations or dynamics, we report costs as mean daily values. Although the costs are expressed in arbitrary units (a.\,u.), they provide interpretable qualitative insights. In our model, a mitigation level of $M = 0.5$ corresponds to a cost of $C_M = 1$\,a.\,u. Due to the superlinear nature of mitigation costs (Fig.~\ref{fig:model}c), increasing mitigation to $M = \frac{2}{3}$ already doubles the cost ($C_M = 2$\,a.\,u). This implies that mitigating a disease with $R_0 = 3$ is approximately twice as costly as mitigating one with $R_0 = 2$. As a result, stronger interventions impose a growing burden, underscoring the need for efficient mitigation strategies. 

\paragraph{(II) Disease dynamics.} We use an SIRS model \cite{SIRS}, with the compartments $S$, $I$, and $R$ representing the susceptible, infectious, and recovered fractions of the population, respectively: 
\begin{align}
\begin{split}
    \dot{S} &= -R_{\textsf{eff}}(t)\cdot \gamma \cdot I\cdot S+\nu\left(1-S-I\right)\\
    \dot{I} &= \:\:\:R_{\textsf{eff}}(t)\cdot \gamma \cdot I\cdot S-\gamma \cdot I\\
    R &= 1-S-I.
\end{split}  
\label{eq:SIRS}
\end{align}
$R_{\textsf{eff}}$ is the effective reproduction number, $\gamma$ the recovery rate, and $\nu$ the waning rate. The effective reproduction number, $R_{\textsf{eff}}$, is modeled as a fractional reduction of the basic reproduction number $R_0$,
\begin{equation}
R_{\textsf{eff}}(t) = R_0(t) \left(1-M(t)\right), 
\label{eq:Reff}
\end{equation}
where $M(t)$ denotes the time-dependent mitigation, and $R_0$ is the basic reproduction number (which can be time-varying because of seasonality). An overview of all parameters is given in table \ref{tab:params}. The initial conditions do not influence the long-term behavior of the SIRS model; regardless of their starting values, the system will eventually reach the same equilibrium or cyclical pattern. Modifications to this model to include different scenarios, like seasonality or vaccination, are described in SI~\ref{SI_methods_implementationscenarios}.

\paragraph{(III) Mitigation and optimization.} The optimal mitigation strategy is determined by minimizing the total cost using an optimization algorithm. We construct the mitigation $M(t)$ by assigning one optimization parameter to each time step ($\hat{=}$ 1\,day), allowing for maximal flexibility. For the optimization, we employ the limited-memory Broyden-Fletcher-Goldfarb-Shanno (L-BFGS) algorithm \cite{liu1989limited}. 
As the number of parameters is on the order of a few hundred, numerical differentiation with finite differences is computationally infeasible. Instead, we use JAX \cite{jax} and Diffrax \cite{kidger2021} in combination with the ICoMo library \cite{Dehning_ICoMo_2024} to integrate the differential equations and compute gradients via automatic differentiation. This approach ensures both efficiency and precision in optimizing the mitigation strategy. 

The framework is visualized in Fig.~\ref{fig:model}a. Code is available at \url{https://github.com/Priesemann-Group/optimizing_pandemic_mitigation}.

\section{Results}

\subsection{Optimal mitigation undergoes a phase transition with disease severity}

In the simplest case of a pandemic with a constant basic reproduction number, $R_0$, the optimal mitigation strategy is, as expected, to keep an approximately steady mitigation level (Fig.~\ref{fig:model}d-f). Transient deviations may occur in the initial phase before equilibrium. Depending on disease severity $a$, the stringency of the mitigation leads to two qualitatively distinct outcomes. 
For severe diseases (e.g., $a = 10, 100 $; Fig.~\ref{fig:model}d,e), high levels of mitigation are optimal, allowing prevalence to decline to very low levels.

In contrast, for mild diseases (e.g. $a=1$, Fig.~\ref{fig:model}f), the cost of mitigation outweighs its benefits, making it more cost-efficient not to implement any mitigation measures at all. 

In the absence of mitigation, we observe equilibration dynamics at the beginning, particularly pronounced for mild diseases, here $a=1$, where mitigation is absent and multiple infection waves occur. These waves arise due to the initially naive population and are a characteristic feature of SIRS dynamics. The nature of these transient dynamics depends on the transmission rate, which in this case is effectively reduced by the mitigation factor ($1-M$). For mild diseases ($a=1$), we see full oscillatory waves before settling into equilibrium. In contrast, for more severe diseases ($a=10,\,100$), the system exhibits a smoother monotonic approach to equilibrium without pronounced waves. In all cases, however, prevalence ultimately converges to its equilibrium level.

Our analytical results confirm the emergence of two distinct regimes depending on disease severity (Fig.~\ref{fig:model}g). As severity increases, a phase transition occurs once a threshold $a_c$ is surpassed (here, $a_c\approx4.4$). For mild diseases ($a<a_c$), the optimal mitigation remains zero, and the prevalence follows the standard SIRS equilibrium. In contrast, for severe diseases ($a>a_c$), optimal mitigation abruptly jumps to $M_{opt} \approx 1-1/R_0 = 0.5$, leading to a sharp drop in the equilibrium prevalence\,—\,from approximately 4.5\,\% to approximately 0.2\,\%\,—\,and soon reaches 0\,\%.
This result highlights a clear principle: in long-term SIRS dynamics, mild diseases require no mitigation, while severe diseases display an optimal mitigation level of $M_{opt} = 1-1/R_0$; intermediate mitigation is inefficient.

In the following analyses, we focus on the more relevant case of severe diseases requiring mitigation, and fix the severity parameter at $a = 10$ unless stated otherwise.

\subsection{Optimal mitigation alters seasonal outbreak severity and timing}

\begin{figure}[hbtp]
    \centering\includegraphics{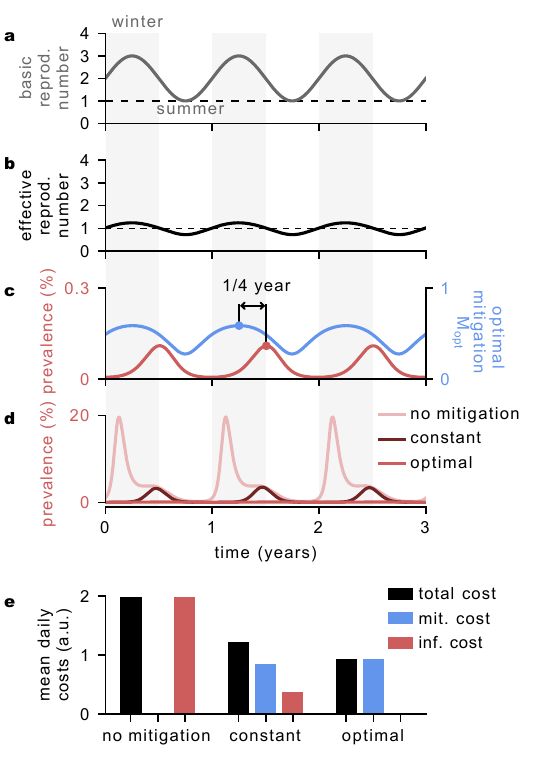}
    \caption{
    \textbf{Seasonally timed mitigation delays and flattens infection waves. }
    \textbf{(a)} We assume a sinusoidal time evolution of the basic reproduction number, representing the seasonality of disease transmission (maximal in winter, minimal in summer). \textbf{(b)} With an optimal mitigation strategy, the resulting effective reproduction number ($R_{\textsf{eff}}(t)=R_0(t)\left(1-M(t)\right)$) has a lower amplitude than the basic reproduction number and oscillates around 1. \textbf{(c)} The peak of the resulting prevalence is shifted by $\pi/2$ (1/4 year) compared to the peak of the optimal mitigation. \textbf{(d)} Compared to scenarios with no mitigation or a comparable constant mitigation, the prevalence of the optimal strategy is significantly lower. \textbf{(e)} In the case of no mitigation, the mean daily costs, which are only comprised of infection costs, are approximately double those of the optimal scenario. In the constant scenario, mitigation costs are similar to those in the optimal scenario, however, infection costs are notably higher, leading to higher total costs. Results for different disease severities, cost functions, mean values, and amplitudes of the basic reproduction number can be found in Supplementary Figs. \ref{fig:SI_seasonality_b} to \ref{fig:SI_seasonality_deltabeta}.}
    \label{fig:seas:general}
\end{figure}

Many infectious diseases, particularly respiratory infections, exhibit seasonal patterns in incidence.
These patterns are often linked to fluctuations in the basic reproduction number, with transmission typically increasing during winter and decreasing in summer, as observed for Influenza virus, human coronavirus (HCoV), and human respiratory syncytial virus (RSV) \cite{Moriyama2020}. We implement these seasonal fluctuations using a sinusoidal basic reproduction number with a period of one year (Fig.~\ref{fig:seas:general}a; Supplementary Eq.~\eqref{eq:seasoanlity}). 

In the presence of seasonal fluctuations, the optimal mitigation also follows an approximately sinusoidal pattern, oscillating in phase with the basic reproduction number (Fig.~\ref{fig:seas:general}c). Thereby, the mitigation measures counteract the oscillations of the basic reproduction number. The resulting effective reproduction number (Eq. \eqref{eq:Reff}) also displays a periodic pattern (Fig.~\ref{fig:seas:general}b). However, it oscillates around 1 and has a lower amplitude compared to the basic reproduction number ($\pm 0.26$ instead of $\pm 1$). 

\begin{figure*}[htb]
    \includegraphics{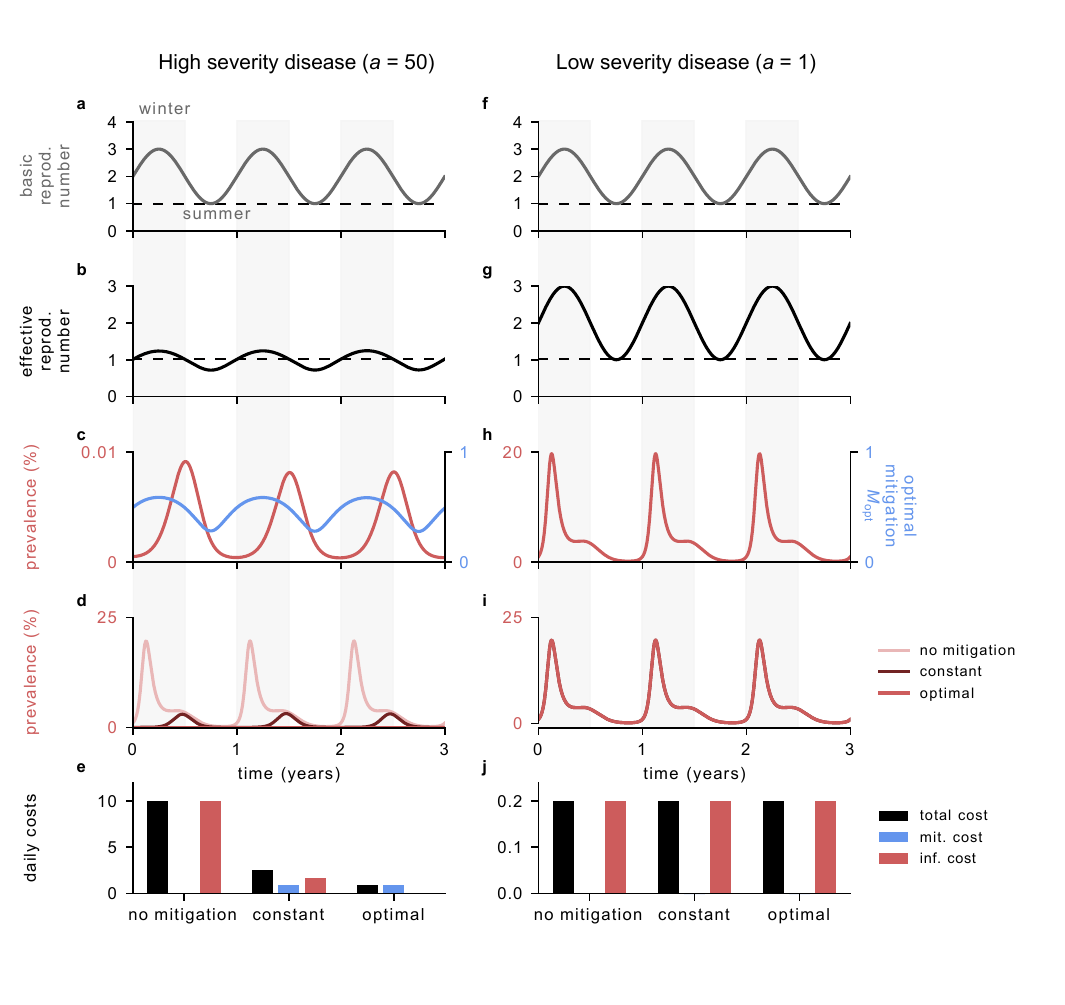}
    \caption{\textbf{Comparison between different disease severities $a$ in the presence of seasonality.} 
    We extend Fig.~\ref{fig:seas:general} ($a = 10$) to examine how disease severity affects optimal mitigation strategies: $a=50$ (panels \textbf{a-e}) represents a severe disease, and $a=1$ (panels \textbf{f-j}) represents a mild disease.
    For the severe disease ($a = 50$), the high cost of infections drives strong mitigation that oscillates in phase with the basic reproduction number, keeping the effective reproduction number near 1 and prevalence low. While this results in minimal seasonal variation in prevalence (panel \textbf{c}), the infection waves are several orders of magnitude smaller than without mitigation.
    For the mild disease ($a = 1$), no mitigation is applied and prevalence is high with pronounced seasonal waves. The low infection cost makes tolerating this high prevalence economically optimal.
    These contrasting outcomes demonstrate how disease severity fundamentally shapes optimal pandemic response: severe diseases require strong, seasonally adapted suppression while mild diseases may tolerate higher infection levels with no  control.}
    \label{fig:SI_seasonality_b}
\end{figure*}

This qualitative behavior is robust; for a wide range of parameters and different cost functions, the optimal mitigation will resemble the shape of the basic reproduction number and maintain the effective reproduction number around one. The exact mitigation strategy, however, depends on the cost function: For example, for a logarithmic mitigation cost (Supplementary Eq.~\eqref{eq:CM_log}), the optimal mitigation oscillates in such a way that the effective reproduction number remains approximately constant at 1 throughout the whole year (Fig.~S\ref{fig:SI_seasonality_CM}). For other cost functions (Eq.~\eqref{eq:CM}; Supplementary Eq.~\eqref{eq:CM_sqrt}) the  effective reproduction number oscillates around 1 as the optimal mitigation is not able to completely balance the basic reproduction number under these costs. For mild diseases, mitigation is absent and infection dynamics are primarily controlled by immunity (Fig.~S\ref{fig:SI_seasonality_b}).

\begin{figure}[H]
    \centering
    \includegraphics{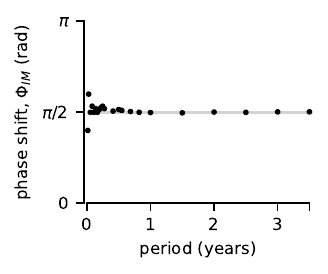}
    \caption{\textbf{The phase shift between the peaks of the mitigation and infection waves does not depend on the period of the seasonality.}
    In Fig.~\ref{fig:seas:general}, we observed that both the mitigation and prevalence have one annual wave. The peak of the mitigation is in winter, whereas the prevalence peak is in spring. This corresponds to a phase-shift of one season, or $\frac{\pi}{2}$, between the two peaks. To determine whether this phase-shift is dependent on the period of the sine curve that we used to model the seasonal fluctuations, we modified the periodicity of the seasonal oscillations of the effective reproduction number, from 5 days to 3 years. For each of these periodicities, we calculated the shift of the mitigation peak compared to the prevalence peak. The phase-shift is always $\frac{\pi}{2}$, indicating that it is independent of the period. The deviations from $\frac{\pi}{2}$ for small values of the periodicity are due to numerical errors induced by our time discretization of 1 day.
    }
    \label{fig:SI_seasonality_phaseshift}
\end{figure}

Interestingly, despite both the basic and effective reproduction numbers being maximal in winter, the peak of the prevalence is in spring (Fig.~\ref{fig:seas:general}c). 
Prevalence increases for half of the year from mid-fall to mid-spring, where the effective reproduction number exceeds 1. 
For the other half of the year, prevalence decreases. Consequently, it peaks in spring, where the effective reproduction number crosses 1.
Between the peaks of mitigation (winter) and prevalence (spring), there is a shift of exactly three months, or $\pi/2$. This shift of $\pi/2$ is independent of the period of the fluctuations (Fig.~S\ref{fig:SI_seasonality_phaseshift}).

To evaluate the benefits of optimal mitigation, we compare it to two alternative strategies: no mitigation ($M(t) = 0$) and constant mitigation set to the mean value of the optimal strategy ($M(t) = \overline{M_{\textsf{opt}}(t)}$) such that the overall mitigation cost is comparable (Fig.~\ref{fig:seas:general}d, e). In the absence of mitigation, a large infection wave occurs in winter, followed by a period of sustained prevalence in spring (Fig.~\ref{fig:seas:general}d). 
Under constant mitigation (with the same mean value of mitigation as in the optimal case), a single but lower wave peaks in early spring (reduced by 82\,\%), reducing the total cost by 38\,\% compared to no mitigation (Fig.~\ref{fig:seas:general}e). 
Optimal mitigation also results in a single peak in prevalence in spring; however, its height is considerably lower---by 99\,\% compared to no mitigation and by 97\,\% compared to constant mitigation---rendering it barely visible in Fig.~\ref{fig:seas:general}d. The optimal total cost decreases by 53\,\% compared to the no mitigation scenario and by 24\,\% compared to the constant mitigation, with infection costs nearly absent (Fig.~\ref{fig:seas:general}e). 
This comparison highlights a key policy insight: Not only mean mitigation but also its precise temporal profile matters; even with identical mean mitigation and nearly identical mitigation costs, the total costs and infection burdens can differ substantially.

In summary, optimal mitigation in the presence of seasonality not only reduces the total costs and outbreak severity but also shifts the seasonal peak from winter to spring (in the model precisely by $\pi/2$). The reduction in total costs underscores the value of proactive mitigation strategies implemented ahead of anticipated waves.

\subsection{Vaccination allows for a decrease in mitigation measures}

Vaccinations, when they become available, play a crucial role in disease management. They decrease disease transmission within a population for a sustained period. Even if vaccination does not guarantee complete immunity, people are less likely to get infected or infect others if they are vaccinated \cite{HALL2021,LevineTiefenbrun2021}.

\begin{figure}[hbtp]
    \centering\includegraphics{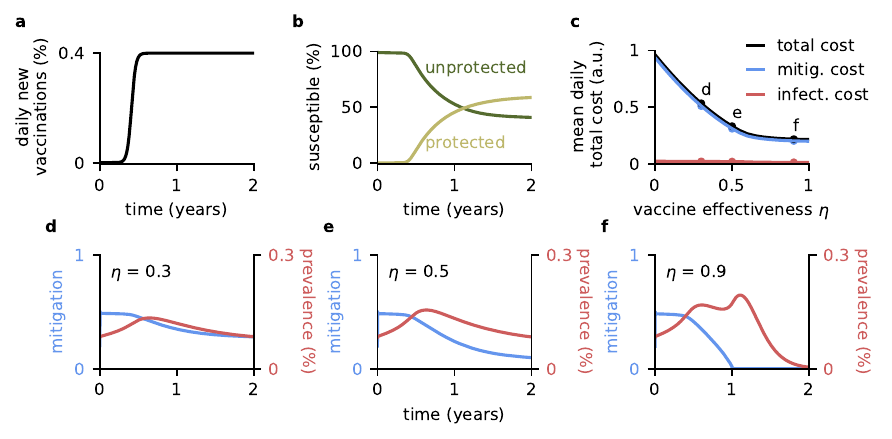}
    \caption{ \textbf{Impact of vaccine rollout on optimal mitigation and prevalence patterns.}
    \textbf{(a)} We assume a rapid increase of the daily new vaccinations at the beginning of the vaccination campaign before they plateau. This steady distribution mainly represents booster vaccinations keeping a part of the population partially immune (Eq.~\eqref{eq:vacc_Nvacc}). \textbf{(b)} 
    The whole population will not reach the vaccinated (protected) status as we assume a waning vaccine immunity ($\nu_\textsf{vac} = 1/150$ day$^{-1}$, see Supplementary Eq.~\eqref{eq:vacc_S}, \eqref{eq:vacc_Svacc}). \textbf{(c)} The mean daily total costs decrease with increasing vaccine effectiveness up to the point where the vaccine is effective enough to eradicate the disease on its own. The values for the example plots in panels d-f are marked with dots. The cost is integrated over the total simulation length (3 years). \textbf{(d-f)} Vaccination enables a reduction in mitigation stringency and, with high vaccine effectiveness, may even allow for a complete halt (f). During this period, however, one or more small infection waves may still occur. Results for different disease severities and cost functions can be found in Supplementary Figs. \ref{fig:SI_vacc_b} and \ref{fig:SI_vacc_CM}.}
    \label{fig:vacc}
\end{figure}

To include the protective effect of vaccination in the SIRS model, we split the \textit{S}, \textit{I} and \textit{R} compartments into two compartments each, one for \textit{vaccinated} and one for \textit{unvaccinated} people (Supplementary Eq.~\eqref{eq:vacc_S}-\eqref{eq:vacc_Rvacc}). Vaccinated individuals are assumed to be less likely to become infected and to transmit the disease. Their contribution to transmission is reduced by a factor $\eta$, which captures the vaccine's effectiveness in lowering overall transmission. Accordingly, we define an adjusted effective reproduction number, $R_\textsf{vac}$, for vaccinated individuals.
\begin{equation*}
    R_\textsf{vac} = R_{\textsf{eff}}(1-\eta).
\end{equation*}
Each day, a fixed number of individuals move from the unvaccinated susceptible compartment to the vaccinated susceptible compartment (Fig.~\ref{fig:vacc}a,b; Supplementary Eq.~\eqref{eq:vacc_Nvacc}). Immunity from vaccination wanes at rate $\nu_{\textsf{vac}}$, returning individuals to the unvaccinated susceptible pool. As starting condition, we assume that the vaccination campaign only starts after a longer time, i.e.,  after the system has reached its endemic equilibrium under optimal mitigation.

The increasing immunity due to vaccination enables a gradual reduction in other mitigation measures (Fig.~\ref{fig:vacc}d--f).
Before the vaccine rollout begins, the optimal mitigation strategy is approximately constant, similar to the baseline scenario without vaccination (Fig.~\ref{fig:model}d-f). As vaccination progresses, the number of unvaccinated susceptible people declines (Fig.~\ref{fig:vacc}b). The extent to which mitigation then decreases depends on the vaccine effectiveness $\eta$. For sufficiently high vaccine effectiveness, mitigation measures can even be stopped entirely (Fig.~\ref{fig:vacc}f). 
Interestingly, even under optimal control, small infection waves may occur during the early stages of vaccine rollout, before the population reaches a new equilibrium (Fig.~\ref{fig:vacc}d-f). This highlights that optimal mitigation strategies during vaccination campaigns may involve non-trivial time dynamics and can yield counterintuitive patterns in infection prevalence. 

The resulting total cost decreases non-linearly with increasing vaccine effectiveness (Fig.~\ref{fig:vacc}c). A vaccine effectiveness of $\eta = 0.5$ already reduces the total cost by 65\,\% compared to the scenario without vaccination. Overall, effective vaccination reduces the need for other measures, leading to substantial economic savings and greater personal freedom.

We note that in real-world scenarios such as the COVID-19 pandemic, vaccination not only reduced transmission, but also significantly lowered disease severity in breakthrough infections \cite{peter2022sars}. In our framework, such effects could be captured by allowing the severity parameter \( a \) to decrease with cumulative vaccination coverage. Note that we do not include vaccination costs in our model since the vaccination rate is independent of the mitigation strategy and we do not aim to optimize vaccine distribution. However, this framework could be expanded to include the cost of the vaccination rate, hence obtaining an estimation of the most cost-effective timing of a vaccination campaign.

\subsection{Delaying mitigation measures entails additional costs}

\begin{figure}[hbtp]
    \centering\includegraphics{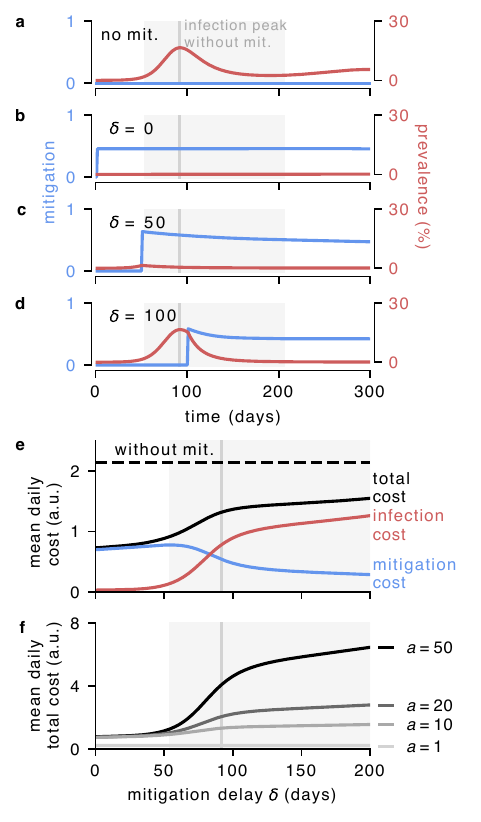}
    \caption{\textbf{Delaying mitigation increases total cost --- especially for more severe diseases. }
    \textbf{(a-d)} Delaying mitigation for (c) 50 or (d) 100 days significantly increases the total prevalence and requires temporary stronger mitigation to control case numbers, which are in all cases lower than without mitigation (a). 
    The gray line and area mark the peak and interval of the first infection wave in the case of no mitigation.
    \textbf{(e)} Total cost increases with delay, driven by higher infection burden—even though mitigation becomes shorter and cheaper. The total simulation length, over which the costs are calculated, is 455 days.
    \textbf{(f)} The change in total costs by delaying mitigation, i.e.~the ``cost of waiting", increases the more severe the disease is. 
    Results for different disease severities and cost functions can be found in Supplementary Figs. \ref{fig:SI_costofwaiting_b} and \ref{fig:SI_costofwaiting_CM}.}
    \label{fig:costofwaiting}
\end{figure}

It is known that delayed onset of mitigation in the case of a new pandemic can incur high additional costs \cite{kompas2021health, pak2020economic, contreras2021}.
Here, we \textit{quantify} these costs by modeling scenarios in which mitigation is initially absent and subsequently applied optimally after a fixed delay~$\delta$, measured from day zero, with $I(0) = 10^{-4}$ (approx. 8{,}000 individuals in Germany), a plausible early prevalence for a country that is not the initial epicenter, given the potential for rapid undetected spread (as observed, for example, in COVID-19).
Specifically, we set \( M(t) = 0 \) for \( t < \delta \) and optimize mitigation for \( t \geq \delta \) (Fig.~\ref{fig:costofwaiting}). For our standard disease severity (\( a = 10 \)), mitigation begins with a sharp increase in intensity and then slowly converges to its equilibrium value, consistent with the non-delayed results (Fig.~\ref{fig:model}d-f). 
In addition, we explore how the results depend on the choice of cost function and find qualitatively similar behavior across variations (Supplementary Fig.~\ref{fig:SI_costofwaiting_CM}).

Prior to the onset of mitigation (\( t < \delta \)), the system follows standard SIRS dynamics (Fig.~\ref{fig:costofwaiting}a–d). The size of the resulting initial infection wave increases with the delay, driving up the total cost (Fig.~\ref{fig:costofwaiting}e,f). While the infection cost consistently increases, the mitigation cost decreases with $\delta$ (apart from a small initial rise) but does not offset the increasing infection cost. The initial rise of the mitigation cost reflects a catch-up effect: the system must mitigate more aggressively once intervention begins in order to counteract an ongoing infection wave. 
Consequently, total cost increases monotonically with delay, though it always remains below the cost of an uncontrolled epidemic. This cost increase becomes more pronounced as disease severity $a$ rises (Fig.~\ref{fig:costofwaiting}f). This finding underscores that even late intervention is still beneficial.

For interventions that start after the peak of the initial infection wave (gray vertical line in Fig.~\ref{fig:costofwaiting}), the total cost increases at a slower rate. This suggests that rapid response is most critical before the first infection peak would occur, as early intervention reduces the prevalence and, consequentially, the total costs of the outbreak (Fig.~\ref{fig:costofwaiting}e). 

To contextualize the cost of delay, consider the case of \( a = 10 \). Even a short delay of $\delta=7$ days results in 0.39\,\% more of the population getting infected during the simulation compared to immediate intervention, i.e. $\delta=0$ (5.55\,\% vs. 5.16\,\%). In a country like Germany, this translates to roughly 300{,}000 additional infections---highlighting the real-world consequences of even modest delays in implementing mitigation measures. 

Importantly, our framework also allows to consider the total cost, though in arbitrary units. For the example above, waiting 7 days increases the total cost by 1.18\,\%, compared to an increase in infections of 7.58\,\%.
This demonstrates the difficulty in assessing the quality of mitigation strategies using infection numbers alone.

\section{Discussion} 

We developed a computationally efficient framework for optimizing time-dependent mitigation strategies for pandemics and epidemics. Moreover, its versatility allows for adaptation to various infectious diseases, scenarios, and policy objectives by modifying the underlying disease model, parameters, and cost functions.   
For example, incorporating seasonal fluctuations required only minor modifications to the base SIRS model. This makes the framework well-suited to determine effective mitigation strategies during future disease outbreaks. 
We applied the framework to diverse scenarios, including seasonality, vaccination rollout, and delayed intervention, showing several unexpected results.


One of the most striking findings is that optimal mitigation exhibits a sharp transition depending on disease severity. For mild diseases, the cost of intervention outweighs its benefits, and the optimal strategy is to implement \textit{no mitigation at all}. For severe diseases, the optimal strategy abruptly switches to full mitigation at a level \( M = 1 - \frac{1}{R_0} \), which keeps \( R_{\text{eff}} = 1 \). 
There is no stable intermediate regime. This "all-or-nothing" response offers a clear guiding principle: in long-term epidemic control, the choice is binary --- either suppress fully or not at all.


Another important finding is that if seasonal fluctuations in the basic reproduction number are introduced, optimal mitigation is in phase with the seasonality. This finding is consistent with other results \cite{Shirin2021}, which showed that adjusting mitigation to maintain \( R_{\text{eff}} \approx 1 \) is optimal in settings with a time-varying basic reproduction number. As a result, the infection peak is moved from fall/winter to spring, when the probability of getting infected is naturally lower.
This contrasts with real-world responses during COVID-19, where interventions at times reacted to current prevalence rather than anticipating future risk. Such reactive strategies can be explained by communication challenges and behavioral fatigue, which complicate the implementation of preemptive measures \cite{skovdal2020complexities, maguire2022risk, heydari2021effect, rahman2021challenges}.

During vaccination rollout, the optimal strategy involves gradually reducing mitigation as population-level immunity builds. This aligns with previous studies \cite{Bauer2021, Shirin2021, Viana2021}. However, we also observed that small infection waves may still occur even under optimal control. This underscores the fact that the relationship between vaccination, mitigation, and disease prevalence is dynamically complex and not necessarily monotonic. Moreover, it suggests that infection dynamics alone are an unreliable indicator of the optimality of mitigation strategies.

At the onset of a new outbreak, delayed mitigation is known to incur substantial costs \cite{kompas2021health, balmford2020cross}.
A key strength of our framework is that it allows us to quantify this "cost of waiting" precisely. We find that the infection burden increases with longer delay, while mitigation costs may transiently rise as the system ``catches up" to suppress an outbreak, before eventually declining.  Quantitatively, the benefit of acting early is highest before the first infection peak would occur, preventing the high peak; and the stronger the disease burden, the higher the difference between acting earlier or later. Other studies similarly highlight the importance of early implementation and the increased stringency required when measures are delayed \cite{Peni2020,Perkins2020,Albi2021}.

The choices of cost functions are informed by general assumptions about the dynamics of disease control. For mitigation, we assume that costs grow superlinearly with intensity, based on the idea that low-cost measures are implemented first, and later, stricter interventions (e.g., lockdowns) incur successively higher costs. Any rational mitigation strategy follows such a superlinear cost function. This assumption also reflects that instantaneous suppression  (i.e. $R_\textsf{eff}=0$) is practically impossible, due to essential interactions and imperfect compliance. The precise choice of the cost function does not impact the results qualitatively (Fig.~\ref{fig:SI_seasonality_CM}, \ref{fig:SI_vacc_CM}, \ref{fig:SI_costofwaiting_CM}). A limitation of our mitigation cost function is that it is independent of the current prevalence, whereas in real life mitigation is typically cheaper when case numbers are low, e.g. because of test-trace-and-isolate strategies \cite{contreras2021}. Nevertheless, we believe that our choice of an prevalence-independent mitigation cost function is reasonable and general.  

Similarly, for the infection costs, we assume a function that increases superlinearly with prevalence up to the infection threshold $I \approx 0.1\,\%$, capturing the increasing stress to the healthcare system. This choice of infection cost function with a gradual increase of the per-infection cost reflects the increasing mortality rates from inadequate care, and that overwhelmed healthcare systems lead to increased treatment costs, longer recovery times, and higher indirect costs such as long-term health complications and economic disruptions. While exact cost curves are hard to quantify empirically, our assumptions are grounded in general behavioral and systemic considerations, and (akin to the cost of mitigation), the qualitative results we report remain consistent under different functional forms (Supplementary Sec.~\ref{SI_additionalplots}). 
Finally, our framework allows users to easily modify the underlying functions with minimal effort or technical expertise. To illustrate this flexibility, we provide a publicly accessible Google Colab notebook showcasing one scenario from our analysis at \url{https://colab.research.google.com/drive/16J2hNb519R0QZyRJm_egVJGH-J5Lbjfh?usp=sharing}.

The disease severity parameter $a$ determines the ratio between infection and mitigation cost. It captures two distinct aspects. First, it reflects the intrinsic severity of the disease --- such as the risk of hospitalization, long-term complications, or death --- which determines the direct burden of infection. Second, it encodes normative or cultural judgments about how infection-related harms should be weighed against the societal and economic costs of mitigation. This includes, for example, how a population values the prevention of illness or loss of life relative to personal freedom, mental health, or economic stability. Consequently, its value may vary across populations, political settings, and stages of a pandemic. Rather than fixating on a single value, our contribution is to characterize how optimal strategies vary across disease severity regimes and to identify structural principles that hold across wide parameter ranges. It then remains a task for society and politics to understand whether a specific new disease is ``mild" (low $a$) or ``severe", and whether mitigation is warranted. 

In our scenarios, we chose a disease severity of $a=10$ as the default value. This represents a moderately severe disease: It places the system in the mitigation-dominated regime and leads to infection levels that remain below the infection threshold $I \approx 0.1\,\%$, or 100 per 100,000 people; Fig.~\ref{fig:model}), which reflects approximately the hospitalization capacity for COVID-19 in many countries. Hence, effectively, a severity of $a=10$ represents an infectious disease like COVID-19 before immunization.


There are several extensions one could implement to make the model more realistic. For instance, while we implemented seasonality as a sinusoidal fluctuation in \( R_0 \), real-world seasonal patterns are often more irregular and differ across years. This could lead to non-optimal mitigation, simply because society makes wrong assumptions about timing and amplitude of the seasonal changes. Even without seasonality, in reality, the basic reproduction number has variations that are characterized by fluctuations that may be only partially predictable \cite{Sherratt2023},
whereas our study assumes perfect knowledge and deterministic dynamics. To simulate such scenarios, techniques like model predictive control would be necessary \cite{kalman1960, selley2015}.

The mitigation variable in our model does not correspond to a specific intervention, but rather to its overall effect on transmission.  
This abstraction allows our results to be interpreted flexibly: different combinations of measures (e.g., mask use, social distancing, ventilation) can be used to achieve a given mitigation level. We also do not define whether the intervention is mandatory or voluntary. 
The key insight is that optimal mitigation often counteracts the basic reproduction number, rather than reacting to current prevalence --- a principle that can guide the design of proactive and cost-efficient responses in practice.
Thus, although our findings are not intended as direct prescriptive policy, they provide important strategic insights into rational epidemic management.

In summary, we introduced a general framework for determining optimal mitigation strategies under dynamic epidemic conditions. We showed that for mild diseases, it is optimal not to intervene, whereas for severe diseases, mitigation should be adjusted dynamically to maintain $R_{\text{eff}} = 1$.
This enables us to maintain low levels of prevalence, thus preserving healthcare capacities, reducing deaths, and limiting societal disruption \cite{PRIESEMANN2021}. These findings underscore the need for continuous monitoring of transmission dynamics and the timely adaptation of mitigation strategies.

\section*{Code availability}
We provide the code for the general framework, scenarios, and generating graphics at \url{https://github.com/Priesemann-Group/optimizing_pandemic_mitigation}.

\section*{Acknowledgments}

LM, JD, FS, and VP received support from the Max Planck Society. 
LM, FS and VP were funded or received funding by the German Federal Ministry for Education and Research for the infoXpand project (031L0300A). VP received funding by the RESPINOW project (031L0298), and from the Deutsche Forschungsgemeinschaft (DFG, German Research Foundation) under Germany’s Excellence Strategy-EXC 2067/1-390729940.  JD and VP received funding by the SFB 1528 -- Cognition of Interaction. MFE acknowledges support from the Joachim Herz foundation and Fundación La Caixa (Grant number: LCF/BQ/PI23/11970039).

\paragraph{Author contributions}
Conceptualization: FS, LM, VP
Methodology: LM, FS, JD, MFE
Software: LM, JD
Validation: LM, FS, JD
Formal analysis: LM, FS, JD, MFE, VP
Investigation: LM
Resources: VP
Data Curation: LM
Writing - Original Draft: LM
Writing - Review \& Editing: LM, FS, JD, MFE, VP
Visualization: LM
Supervision: FS, JD, VP
Project administration: LM, FS
Funding acquisition: VP

\section*{Competing interests}
All authors declare that they have no conflicts of interest.

\printbibliography

\newpage

\section{Supplementary material}
\subsection{Extended version of theory and methods}
\subsubsection{Implementation of the different scenarios}
\label{SI_methods_implementationscenarios}
For the implementation of the disease dynamics, we use an SIRS model, which is described by Eq.~\eqref{SI_eq:SIRS}, where $\beta$ is the effective transmission rate, $\gamma$ the recovery rate, and $\nu$ the waning rate. The mitigation $M$ is included in the effective transmission rate: $\beta(t) = \beta_0(t)(1-M(t))$.\\
\begin{align}
\begin{split}
        \dot{S} &= -\beta(t)IS+\nu (1-S-I)\\
        \dot{I} &= +\beta(t)IS-\gamma I
\end{split}
\label{SI_eq:SIRS}
\end{align}

The basic and effective reproduction numbers can be calculated as $R_0(t)=\beta_0(t)/\gamma$ and $R_{\textsf{eff}}(t)=R_0(t)(1-M(t))$, respectively.

The different scenarios are simulated by modifying the basic SIRS model. For the implementation of \textbf{seasonality}, we use a sinusoidal transmission rate (Eq.~\eqref{eq:seasoanlity}), which oscillates around the mean value $\beta_{0,\textsf{mean}}$ with an amplitude of $\Delta\beta$. The period is set to one year, so that the maximum and minimum of the sine function correspond to winter and summer, respectively. 
\begin{align}
    \beta_0(t) =\beta_{0,\textsf{mean}} +\Delta\beta\cdot \sin{\left(2\pi\frac{t}{365}\right)}
\label{eq:seasoanlity}
\end{align}

To simulate \textbf{vaccination}, we split all three compartments into two compartments, each: one for \textit{vaccinated} individuals, denoted by ${\textsf{vac}}$, and one for \textit{unvaccinated} individuals (Eq.~\eqref{eq:vacc_S}-\eqref{eq:vacc_Rvacc}). At a rate of $N_{\textsf{vac}}$ (Eq.~\eqref{eq:vacc_Nvacc}) people are moved from the \textit{unvaccinated} to the \textit{vaccinated susceptible} compartment. This simulates a vaccination campaign. For the vaccination rate we chose a sigmoidal function as in the beginning only few vaccines are available and then an increasing number of people is able to get vaccinated each day. The plateau corresponds to ongoing booster vaccinations. As vaccinated people are less likely to get infected or infect others, we introduce the vaccine effectiveness $\eta$, which decreases the transmission rate if vaccinated people are involved. We assume waning immunity after vaccination, which happens at the rate $\nu_{\textsf{vac}}$.
\begin{equation}
    N_{\textsf{vac}} = \frac{0.004}{1+\exp(-0.1(t-150))}
\label{eq:vacc_Nvacc}
\end{equation}
\begin{align}
    \begin{split}
        dS =& -\beta(t) IS  -\beta(t) (1-\eta)  I_{\textsf{vac}} S + \nu  R\\ & + \nu_{\textsf{vac}} S_{\textsf{vac}} - N_{\textsf{vac}}
    \end{split} \label{eq:vacc_S} \\ 
    \begin{split}
        dS_{\textsf{vac}} =& -\beta(t)(1-\eta) I S_{\textsf{vac}} -\beta(t)(1-\eta)^2 I_{\textsf{vac}} S_{\textsf{vac}} \\&+ \nu R_{\textsf{vac}}- \nu_{\textsf{vac}} S_{\textsf{vac}} + N_{\textsf{vac}} 
    \end{split} \label{eq:vacc_Svacc}\\
    dI =& \beta(t) I S + \beta(t)(1-\eta) I_{\textsf{vac}} S - \gamma I\\
    \begin{split}
        dI_{\textsf{vac}} =& \beta(t)(1-\eta) I S_{\textsf{vac}} + \beta(t)(1-\eta)^2 I_{\textsf{vac}} S_{\textsf{vac}}\\ & - \gamma I_{\textsf{vac}}
    \end{split}\\
    dR =& \gamma I - \nu R\\
    dR_{\textsf{vac}} =& \gamma I_{\textsf{vac}} - \nu R_{\textsf{vac}}
\label{eq:vacc_Rvacc}
\end{align}

\subsubsection{Mitigation}
  
Apart from the different scenarios, the effective transmission rate is influenced by the mitigation (Eq.~\eqref{eq:SIRS}). Mitigation includes both personal and government actions to slow the spread of infection. This strength can be realized by different combinations of measures. It can have value from 0 to 1, where 0 corresponds to no mitigation and 1 to full mitigation, meaning a complete stop of transmission.

As the time evolution of the optimal mitigation is not known, we implement it by setting one parameter for each time step ($\hat{=}$ 1\,day), each of which can have a value between 0 and 1 and is independent of its neighbors. Therefore, the number of parameters needed to model the mitigation equals the simulation length in days, leading to a high-dimensional optimization problem.

\subsubsection{Cost function}

To find the parameters for the optimal mitigation, we need to define a cost function that we want to minimize. This cost function depends on the prevalence as well as the mitigation. The cost function is not expressed in real money values, but is just used for balancing the mitigation and infection costs. The optimal mitigation is defined as the mitigation strategy with the lowest total cost.

\begin{figure}[H]
    \centering
    \includegraphics[]{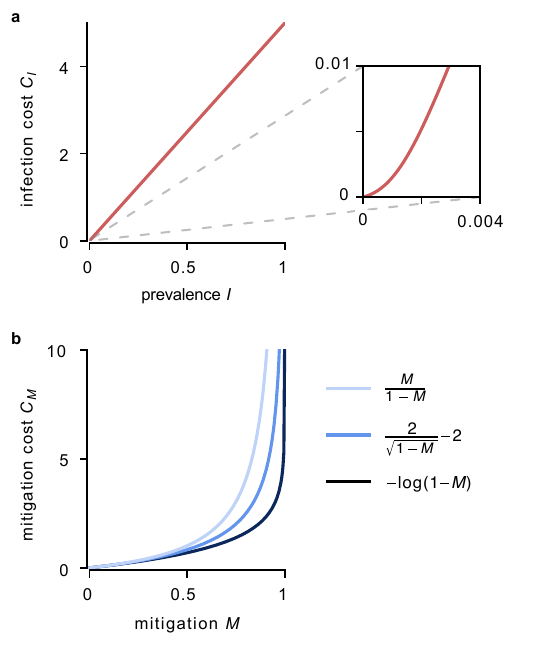}
    \caption{\textbf{Infection and mitigation cost.} 
    \textbf{(a)} The infection cost as a function of prevalence $I$, showing approximately quadratic growth for small prevalence (inset) before transitioning to linear growth. 
    \textbf{(b)} Different mitigation cost functions used in this paper (Eq.~\eqref{eq:CM}, \eqref{eq:CM_log}, \eqref{eq:CM_sqrt}). All functions diverge as $M \rightarrow 1$, reflecting the practical impossibility of complete transmission prevention. The divergence represents that while initial measures (hygiene, masks) are low-cost, achieving higher mitigation levels requires progressively more expensive interventions. The different growth rates illustrate the trade-offs between affordability and achievable mitigation levels: while all functions make extreme mitigation prohibitively costly, less steeply growing functions (darker blue) allow for higher levels of mitigation, while steeper ones (lighter blue) make extreme mitigation prohibitively costly, potentially making infections the cheaper alternative.}
    \label{SI_fig:costplot}
\end{figure}

For the mitigation cost, we use a function that grows rapidly with increasing mitigation (Eq.~\eqref{SI_eq:CM}, Fig.~\ref{SI_fig:costplot}). It diverges, so that a mitigation value of 1 can never be reached, as this is unrealistic in real life. This cost includes social costs, like the mental effects of lockdowns, as well as economic costs, like the actual costs for masks, people not being able to work, etc.
\begin{align}
    C_{M}(t) = \frac{M}{1-M}
\label{SI_eq:CM}
\end{align}

The shape of the infection cost can be split into two parts. It first grows quadratically until the prevalence reaches a value of $I \approx 0.001$, which can be interpreted as a hospitalization limit. Above this threshold, the cost grows linearly with a slope of five, as the healthcare system cannot provide the necessary treatments to everyone anymore. As we need a smooth transition between the two parts for the function to be differentiable, we use Eq.~\eqref{SI_eq:CI} (Fig.~\ref{SI_fig:costplot}). The infection costs, as the mitigation costs, are comprised of social and economic costs. They include hospital costs but also costs for absence from work or even deaths.

\begin{align}
    \begin{split}
        &C_{I}(t)\, =\frac{I}{v}+v\left( I\left(t\right)-0.001\right)\cdot\\ &\quad\quad\quad\left(1-\frac{1}{1+\exp(1000(I\left(t\right)-0.001))}\right)+w, \label{SI_eq:CI}
    \end{split}
\end{align}
where $v = 4.7913$, and $w = 0.0013$.

The total cost consists of the mitigation and infection cost added up and integrated over the whole simulation (Eq.\eqref{SI_eq:Ctot}). The infection cost is multiplied by the disease severity, $a$, which defines the ratio between infection and mitigation cost. For harmless diseases, $a$ is small, which means that the mitigation cost gets a higher relative weight, as we can tolerate high numbers of infectious people and do not need to invest in a lot of mitigation. For more severe diseases, on the other hand, $a$ is high and we get a low weight for the mitigation cost, as we need a lot of mitigation because we do not want to risk lots of infections. 
\begin{align}
    C_{\textsf{tot}} = \int_0^T \left(aC_I(I(t))+ \,C_M(M(t))\right) dt
\label{SI_eq:Ctot}
\end{align}

It is important to note that the choice of the cost function and disease severity may depend on the disease as well as the personal assessment of a person, as everyone weighs the societal impact of the mitigation and seriousness of the prevalence differently.

\begin{figure*}[htb]
    \includegraphics{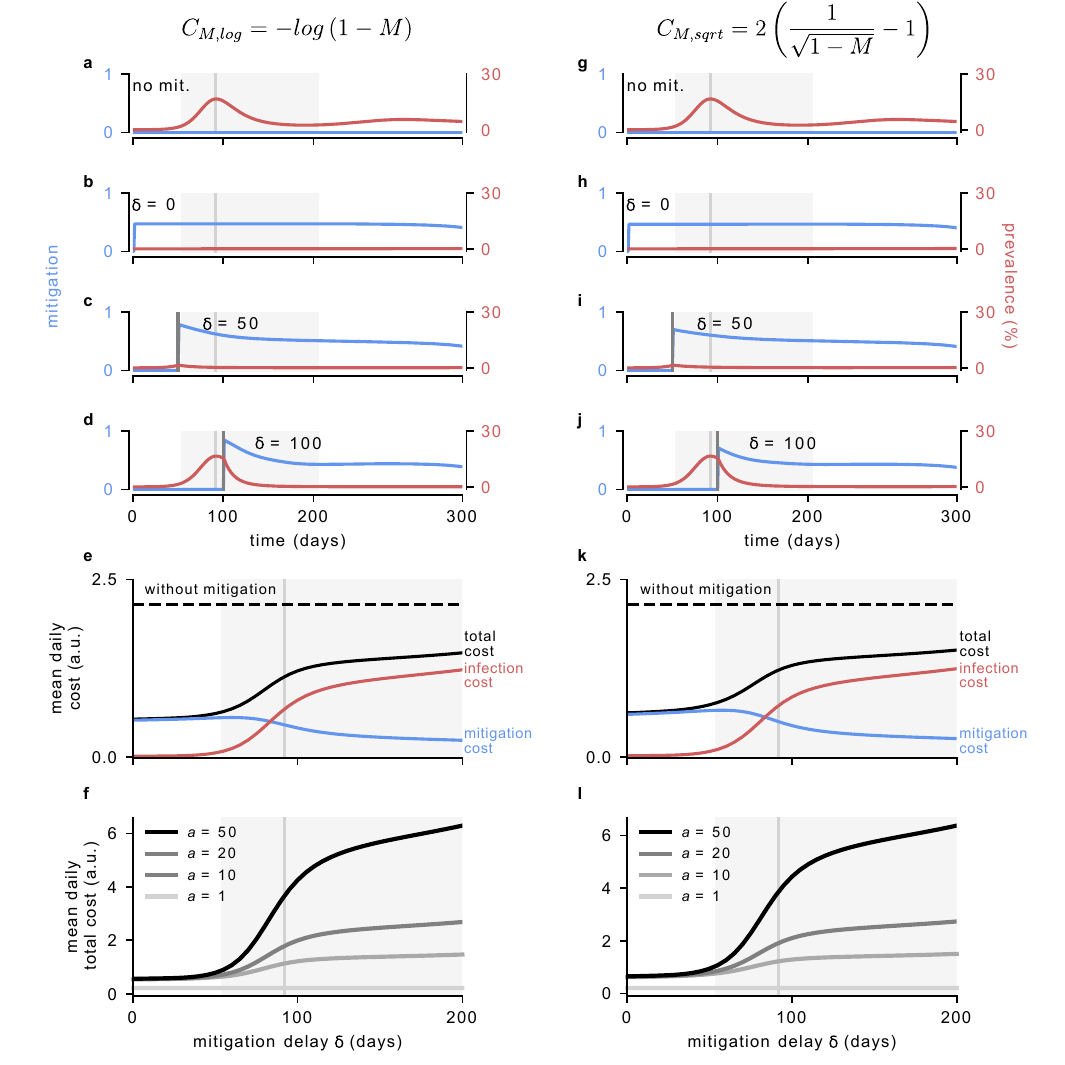}
    \centering
    \caption{\textbf{Comparison between different cost functions in the scenario with delayed mitigation measures.}
    We test how different mitigation cost functions affect the cost of delay using Eq.~\eqref{eq:CM_log} (panels \textbf{a-f}) and Eq.~\eqref{eq:CM_sqrt} (panels \textbf{g-l}), compared to our default in Fig.~\ref{fig:costofwaiting}.
    All cost functions produce qualitatively similar results: mitigation begins at high intensity to control the outbreak, then relaxes. Total costs increase monotonically with delay due to accumulating infections during the unmitigated period.
    While the exact dynamics vary slightly between cost functions, the overall pattern remains robust, demonstrating that our main findings about the cost of delay hold regardless of the specific function chosen for mitigation costs.}
    \label{fig:SI_costofwaiting_CM}
\end{figure*}

To show that the results are qualitatively similar for different cost functions, we carry out the simulations for all scenarios with two additional mitigation cost functions (Eq.~\eqref{eq:CM_log}, \eqref{eq:CM_sqrt}; Fig.~\ref{SI_fig:costplot}). The results can be found in Supplementary Fig.~\ref{fig:SI_seasonality_CM}, \ref{fig:SI_vacc_CM} and \ref{fig:SI_costofwaiting_CM}.

\begin{align}
    \label{eq:CM_log}
    C_{M,\textsf{log}} &= -\log (1-M),\\
    \label{eq:CM_sqrt}
    C_{M,\textsf{sqrt}} &= 2\left(\frac{1}{\sqrt{1-M}}-1\right).
\end{align}

\subsubsection{Parameters}
All parameter information for the SIRS model and scenarios can be found in Tab.~\ref{tab:params}.
\begin{table*}
    \centering
    \begin{tabular}{p{7cm}|p{1.3cm}|p{0.9cm}|p{2.4cm}}
        \textbf{Parameter} & \textbf{Symbol} & \textbf{Unit} & \textbf{Default value} \\
        \hline
        disease severity & $a$ & - & 10 \\
        \hline
        basic reproduction number & $R_0$ & - & 2\\
        \hline
        seasonal fluctations & $\Delta R_0$ & - & 1\\
        \hline
        recovery rate & $\gamma$ & $\text{day}^{-1}$ & 0.1 \\
        \hline
        immunity waning rate after infection & $\nu$ & $\text{day}^{-1}$ & 0.01 \\
        \hline
        immunity waning rate after vaccination & $\nu_\text{vac}$ & $\text{day}^{-1}$ & 1/150 \\
        \hline
        vaccine effectiveness & $\eta$& -& $\in \left[0,1\right]$\\
        \hline
        mitigation delay & $\delta t$ &  days & $\in \left[0,200\right]$\\
    \end{tabular}
    \caption{Default values for parameters. Results for other values can be found in Supplementary Sec. \ref{SI_additionalplots}. As we do not simulate a specific disease in this paper, the values are assumed.  The recovery rate is chosen to be similar to COVID-19 \cite{he_temporal_2020}, and the waning rate of immunity after infection is a little faster than for COVID-19 \cite{goldberg_protection_2022}.}
    \label{tab:params}
\end{table*}

\subsubsection{End of simulation}
\label{SI:sec:endofsim}
\begin{figure}[H]
    \centering
    \includegraphics[]{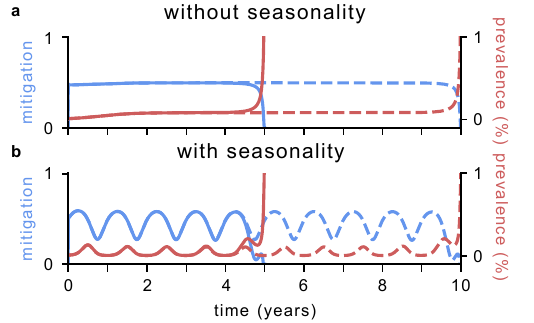}
    \caption{\textbf{Comparison between different simulation lengths.} To show that our results are not artifacts due to the simulation length, we compared two scenarios that differ only in their duration — one lasting 5 years (solid lines) and the other 10 years (dashed lines). The outcomes of both simulations are overlaid for comparison. 
    Panel \textbf{a} displays  mitigation and prevalence for a scenario where the reproduction number is constant, while panel
    \textbf{b} shows the mitigation and prevalence as a function of time in the seasonality scenario.
    In both cases, the results are very similar for the two simulation lengths. Only near the end of each simulation, boundary effects become preponderant, as discussed in Supplementary Sec. \ref{SI:sec:endofsim}.
    }
    \label{fig:SI_general_5vs10}
\end{figure}

To ensure consistent comparison across scenarios, we implement a controlled conclusion to each simulation (i.e. Fig.~\ref{fig:SI_general_5vs10}). We introduce an exponential decrease of the basic reproduction number to 0 over the final 90 days, which can be interpreted as the natural endpoint of epidemic dynamics due to accumulated population immunity.

This approach serves multiple purposes: First, it allows the epidemic to reach a natural conclusion within the simulation time frame rather than artificially truncating ongoing dynamics. Second, it ensures that mitigation strategies across different scenarios converge in a comparable manner, facilitating meaningful comparisons between results.

\subsection{Additional plots}
\label{SI_additionalplots}

\begin{figure*}[htb]
    \includegraphics{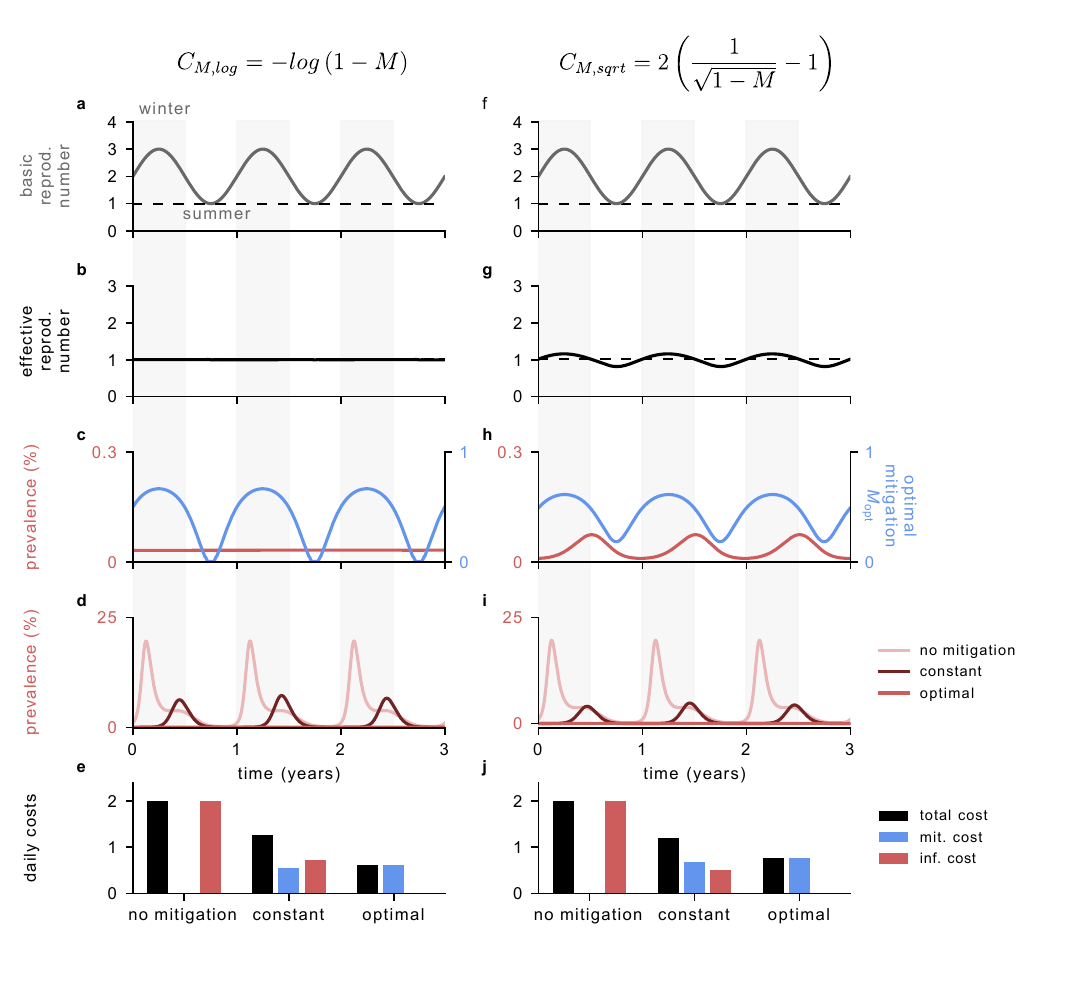}
    \caption{\textbf{Comparison between different cost functions in the presence of seasonality.} 
    We replicate the scenario from Fig.~\ref{fig:seas:general} using two alternative mitigation cost functions: logarithmic (Eq.~\eqref{eq:CM_log}, panels \textbf{a-e}) and square root (Eq.~\eqref{eq:CM_sqrt}, panels \textbf{f-j}).
    With the logarithmic cost function, which grows more slowly than our default (see Fig.~\ref{SI_fig:costplot}), mitigation can afford to be strong enough to completely counteract seasonal fluctuations, maintaining $R_{\text{eff}} \approx 1$ throughout the year. This results in approximately constant, low prevalence without seasonal waves (panels \textbf{a-e}).
    The square root cost function produces intermediate results: the effective reproduction number oscillates around 1, creating annual infection waves peaking in spring, similar to our default cost function but with slightly different dynamics (panels \textbf{f-j}).
    These results demonstrate that while the qualitative strategy (counteracting seasonality with mitigation) remains consistent, the achievable level of control depends on how steeply mitigation costs rise.}
    \label{fig:SI_seasonality_CM}
\end{figure*}

\begin{figure*}[htb]
    \includegraphics{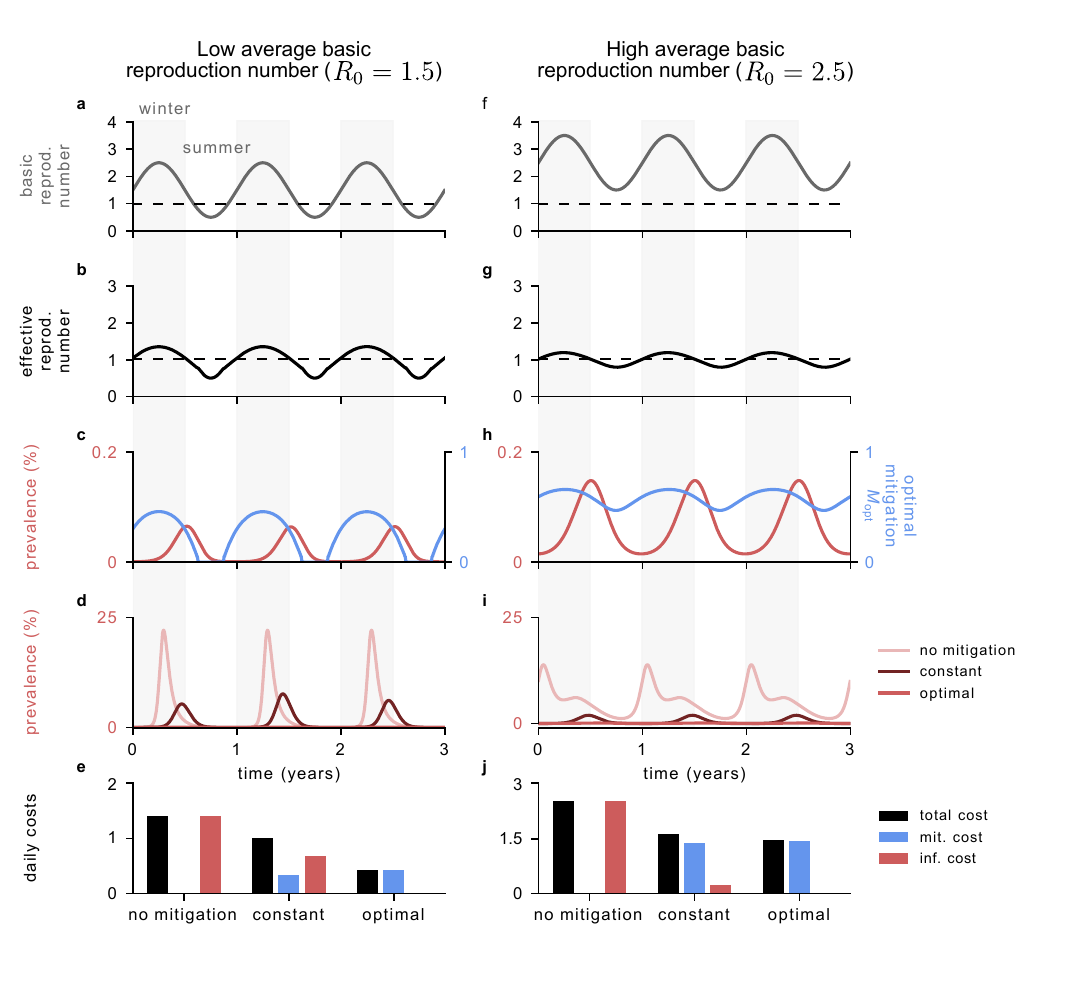}
    \caption{\textbf{Comparison between different mean values $R_{0,\text{mean}}$ of the seasonal fluctuations in the basic reproduction number.} 
    We examine how the mean basic reproduction number affects optimal mitigation using $R_{0,\text{mean}} = 1.5$ (panels \textbf{a-e}) and $R_{0,\text{mean}} = 2.5$ (panels \textbf{f-j}), keeping the seasonal amplitude fixed at 1. For reference, Fig.~\ref{fig:seas:general} shows results for $R_{0,\text{mean}} = 2$.
    For $R_{0,\text{mean}} = 1.5$, the basic reproduction number drops below 1 for approximately one quarter of the year (summer), allowing mitigation to cease entirely during this period. Mitigation peaks in winter when transmission is highest, resulting in a single spring infection wave under optimal control.
    For $R_{0,\text{mean}} = 2.5$, year-round mitigation is required since even during summer, the basic reproduction number is greater than one. The optimal strategy maintains small oscillations in $R_{\text{eff}}$ around 1, producing spring infection waves similar to the default scenario but with higher total costs due to the increased mitigation burden.
    These results show that the baseline transmission level determines whether seasonal "mitigation holidays" are possible (low $R_{0,\text{mean}}$) or whether continuous control is necessary (high $R_{0,\text{mean}}$).}
    \label{fig:SI_seasonality_beta0}
\end{figure*}

\begin{figure*}[htb]
    \includegraphics{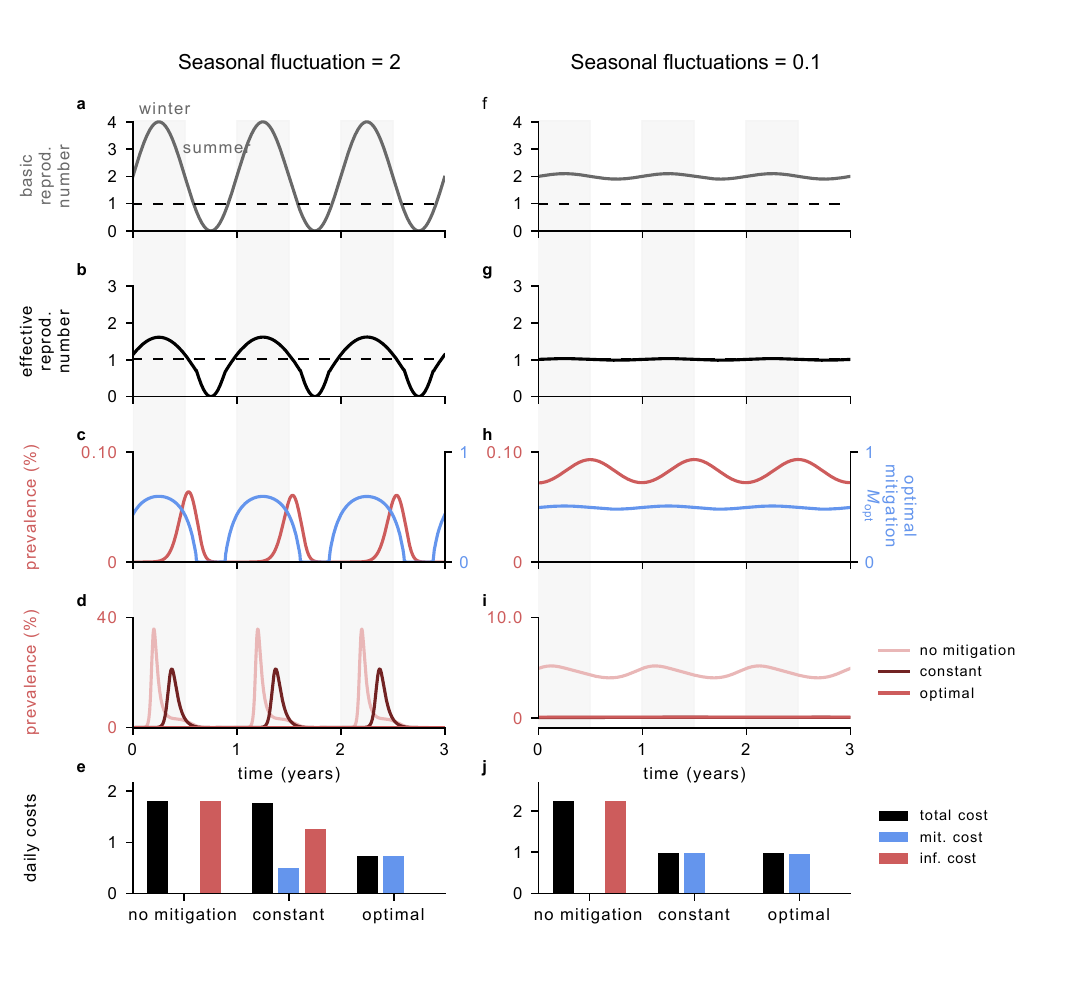}
    \caption{\textbf{Comparison between different amplitudes $\Delta R$ of the seasonal fluctuations in the basic reproduction number.}
    In Fig.~\ref{fig:seas:general}, we had a mean basic reproduction number of 2 with an amplitude of 1. Here, we will keep the mean basic reproduction number of the seasonal fluctuations at 2 while varying the amplitude. We use $\Delta R = 2$ and $\Delta R = 0.1$ (panels \textbf{a-e}, and \textbf{f-j}, respectively).\\
    For an amplitude of 2 (panels \textbf{a-e}), the basic reproduction number falls below 1 for about a quarter of the year, similar to the case above where we varied the mean basic reproduction number. For the optimal mitigation we get similar results: when the basic reproduction number is below one, the mitigation is zero. Otherwise, it increases, with a peak in winter.
    For an amplitude of 0.1 (panels \textbf{f-j}), the mitigation is nearly constant, as the oscillations of the basic reproduction number are small compared to its mean value (5\,\%). Consequently, the prevalence exhibits only small oscillations.
    In conclusion, the amplitude of the seasonality significantly influences the shape of the optimal mitigation, which ranges from nearly constant (low amplitude) to sinusoidal (medium amplitude) to partly zero throughout the year (high amplitude).}
    \label{fig:SI_seasonality_deltabeta}
\end{figure*}

\begin{figure*}[htb]
    \includegraphics{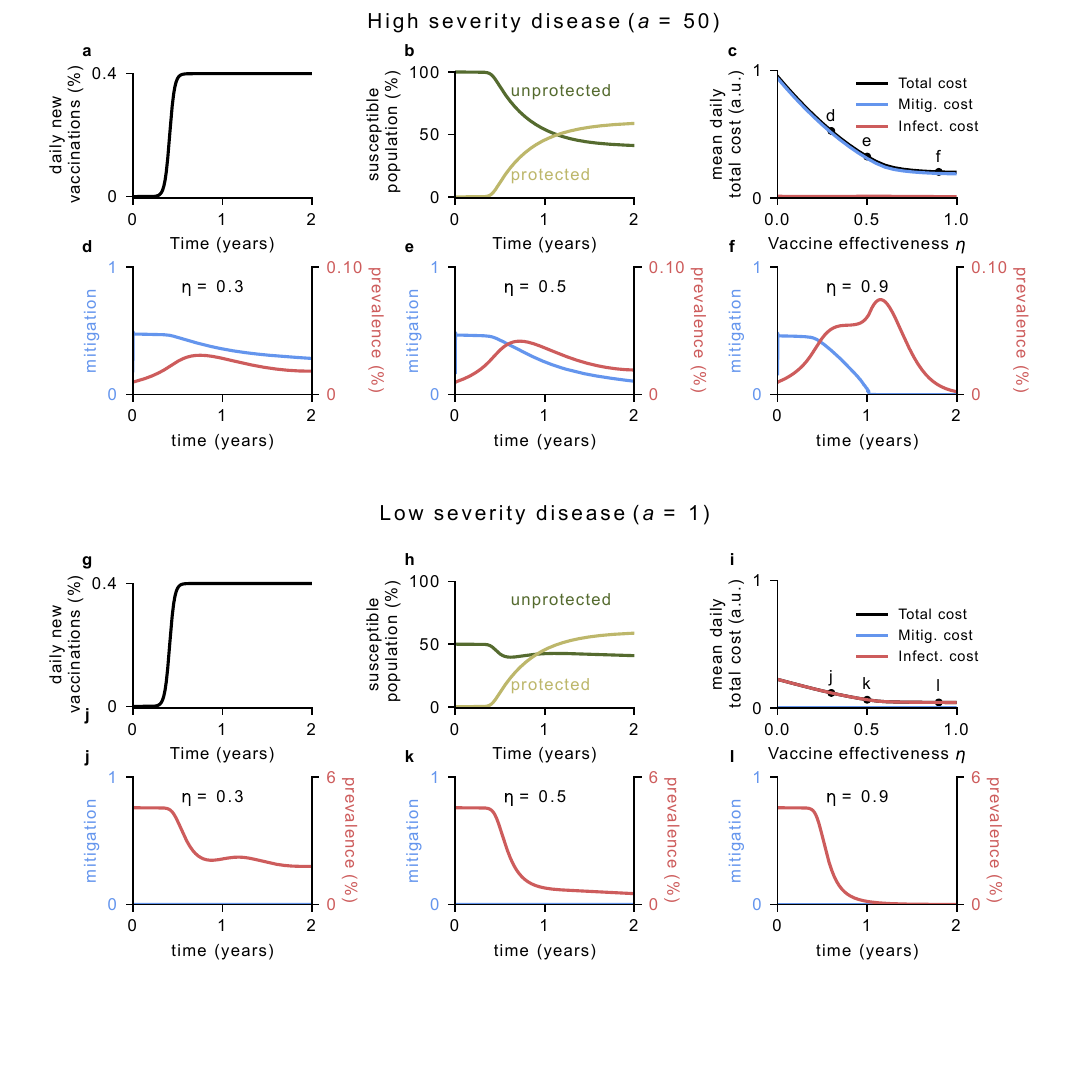}
    \caption{\textbf{Comparison between different disease severities $a$ in the vaccination scenario.}
    In Fig.~\ref{fig:vacc}, we simulated a vaccination campaign with disease severity $a=10$. Here, we show how the same vaccination scenario for two different severities, $a=50$ and $a=1$ (panels \textbf{a-f} and \textbf{g-l}, respectively), while keeping the other parameters fixed as in Fig.~\ref{fig:vacc}. 
    For $a = 50$ (panels \textbf{a-f}), the high infection cost requires strong mitigation throughout the vaccination campaign, keeping prevalence minimal. Therefore, the total cost is dominated by mitigation rather than infection costs.
    For $a = 1$, (panels \textbf{g-l}), the low infection cost makes mitigation uneconomical, so it remains at zero throughout. Here, total costs are dominated by infection costs, in contrast to the higher severity scenarios.
    These results align with the results for different disease severities in the other scenarios: for higher severity we get higher mitigation values and lower prevalence, and for less severe diseases we observe that no vaccination is optimal.
    }
    \label{fig:SI_vacc_b}
\end{figure*}

\begin{figure*}[htb]
    \centering
    \includegraphics{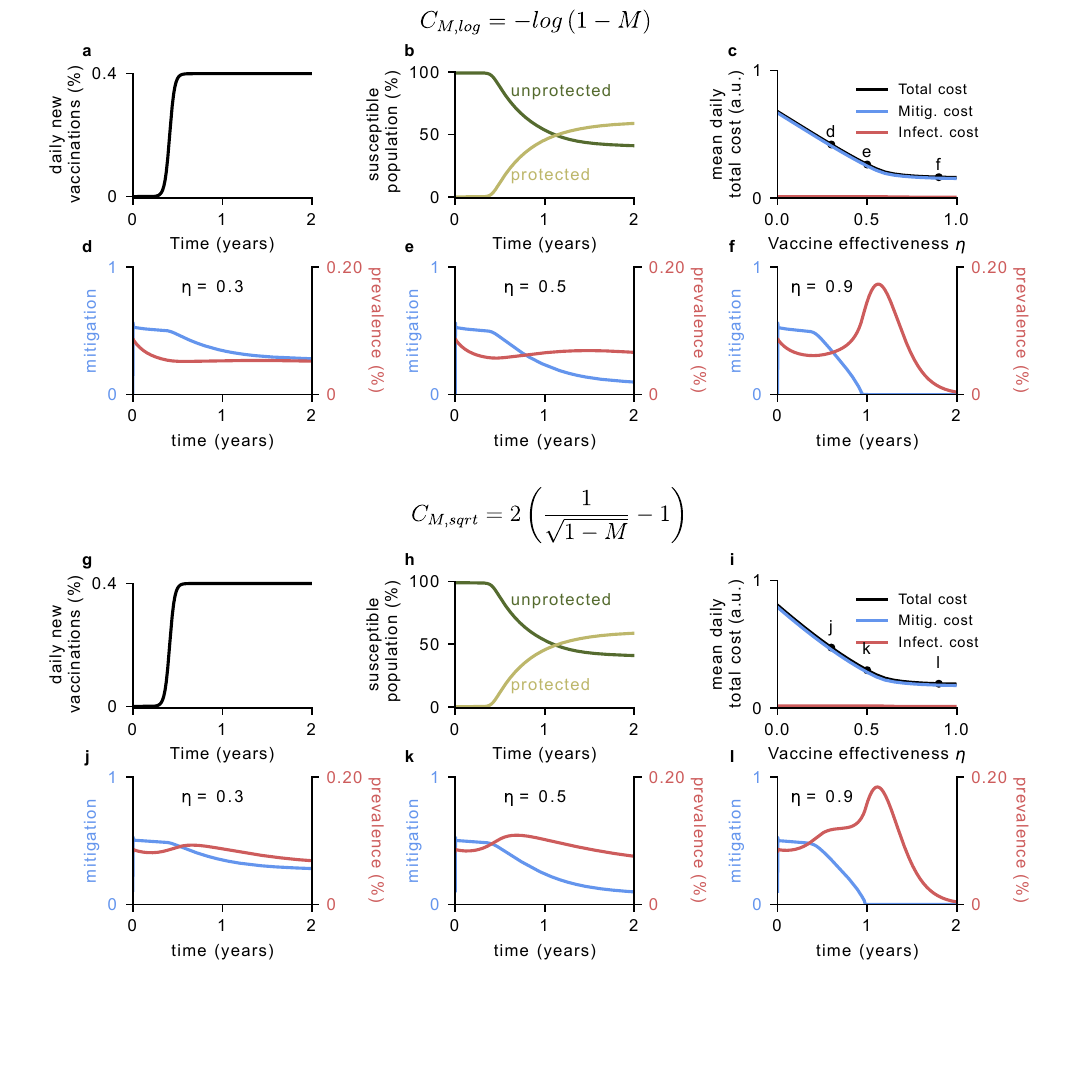}
    \caption{\textbf{Comparison between different cost functions in the vaccination scenario.}
    We test the robustness of our vaccination results (Fig.~\ref{fig:vacc}) using alternative mitigation cost functions: Eq.~\eqref{eq:CM_log} (panels \textbf{a-f}) and Eq.~\eqref{eq:CM_sqrt} (panels \textbf{g-l}), with all other parameters unchanged. 
    The qualitative behavior remains consistent across all cost functions: mitigation decreases during the vaccination campaign, with greater reductions for more effective vaccines. However, the precise dynamics differ between cost functions, leading to variations in prevalence patterns and the timing of transient infection waves.}
    \label{fig:SI_vacc_CM}
\end{figure*}

\begin{figure*}[htb]
    \centering
    \includegraphics{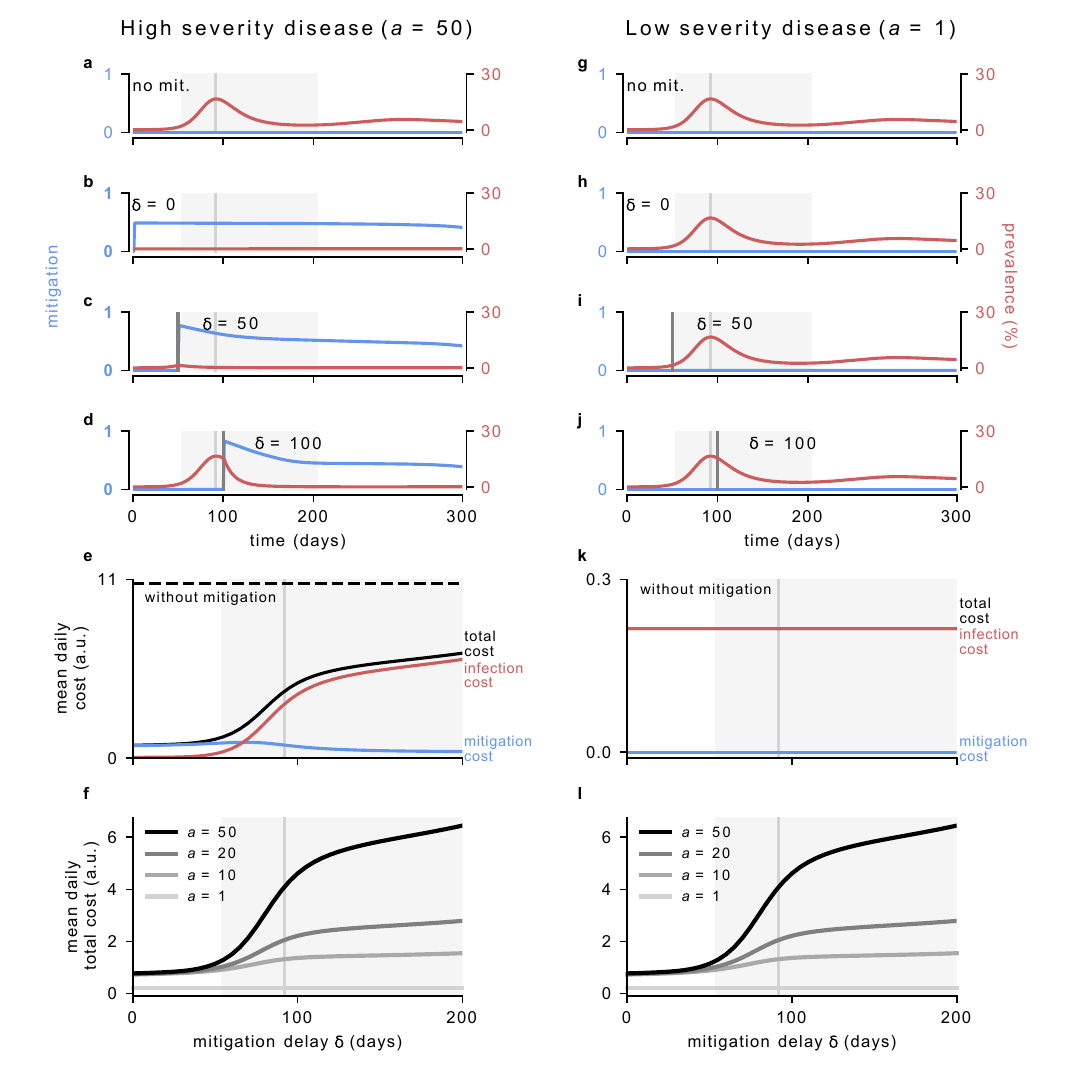}
    \caption{\textbf{Comparison between different disease severities $a$ in the scenario with delayed mitigation measures.}
    We examine how disease severity affects the cost of delayed intervention by comparing $a=50$ (panels \textbf{a-f}) and $a=1$ (panels \textbf{g-l}) with our default $a=10$ shown in Fig.~\ref{fig:costofwaiting}.
    For severe disease ($a=50$, panels \textbf{a-f}), delayed mitigation begins at high intensity to rapidly suppress the ongoing outbreak, then relaxes to maintain low equilibrium prevalence. Total costs increase substantially with delay due to the accumulation of infections during the unmitigated period.
    For mild disease ($a=1$, panels \textbf{g-l}), optimal mitigation remains at zero regardless of delay, as the low infection cost makes intervention uneconomical. Consequently, total costs remain constant with delay and are dominated entirely by infection costs.
    These results demonstrate that the cost of delay is most pronounced for severe diseases, while mild diseases show no penalty for delayed response since intervention is never optimal.}
    \label{fig:SI_costofwaiting_b}
\end{figure*}

\end{document}